\documentclass[
reprint,
aps,
twocolumn,
superscriptaddress,
longbibliography]{revtex4-1}

\usepackage{bm}
\usepackage{amssymb}
\usepackage{amsfonts}
\usepackage{amsmath}
\usepackage{graphicx}
\usepackage{theorem}
\usepackage{color}
\usepackage{soul}
\usepackage{todonotes}
\usepackage[normalem]{ulem}

\usepackage{hyperref}
\usepackage{xcolor}
\hypersetup{
	colorlinks,
	linkcolor={blue!50!black},
	citecolor={blue!50!black},
	urlcolor={blue!80!black}
}

\newcommand{\us}{\mathbf{u}}
\newcommand{\N}{\mathcal{N}}
\newcommand{\M}{\mathcal{M}}
\newcommand{\brk}[1]{\left(#1\right)}

\newcommand{\dif}[0]{\mathrm{d}}

\newcommand{\figref}[1]{Fig.~\ref{#1}}

\newcommand{\Eqref}[1]{Eq.~\ref{#1}}

\newcommand{\appref}[1]{Appendix~\ref{#1}}

\newcommand{\g}{\mathfrak{g}}
\newcommand{\gbar}{\bar{\mathfrak{g}}}

\newcommand{\A}{\mathcal{A}}

\newcommand{\xvec}{\mathbf{x}}

\renewcommand{\d}{\mathbf{d}}

\newcommand{\bfsigma}{\boldsymbol{\sigma}}
\newcommand{\Deltabar}{\bar{\Delta}}
\newcommand{\diag}{\operatorname{diag}}

\begin{document}

\title{Geometric charges and nonlinear elasticity of soft metamaterials}

\author{Yohai Bar-Sinai} 
\affiliation{School of Engineering and Applied Sciences, Harvard University, Cambridge MA 02138}
\author{Gabriele Librandi}
\affiliation{School of Engineering and Applied Sciences, Harvard University, Cambridge MA 02138}
\author{Katia Bertoldi}
\affiliation{School of Engineering and Applied Sciences, Harvard University, Cambridge MA 02138}
\author{Michael Moshe}
\email{michael.moshe@mail.huji.ac.il}
\affiliation{Racah Institute of Physics, The Hebrew University of Jerusalem, Jerusalem, Israel 91904}

\begin{abstract}
Problems of flexible mechanical metamaterials, and highly deformable porous solids in general, are rich and complex due to nonlinear mechanics and nontrivial geometrical effects. While numeric approaches are successful, analytic tools and conceptual frameworks are largely lacking. Using an analogy with electrostatics, and building on recent developments in a nonlinear geometric formulation of elasticity, we develop a formalism that maps the elastic problem into that of nonlinear interaction of elastic charges.
This approach offers an intuitive conceptual framework, qualitatively explaining the linear response, the onset of mechanical instability and aspects of the post-instability state.  
Apart from intuition, the formalism also quantitatively reproduces full numeric simulations of several prototypical structures.
Possible applications of the tools developed in this work for the study of ordered and disordered porous mechanical metamaterials are discussed.
\end{abstract}
\maketitle

\section{Introduction}

The hallmark of condensed matter physics, as described by P.W. Anderson in his paper ``More is Different'' \cite{Anderson1972}, is the emergence of collective phenomena out of well understood simple interactions between material elements. 
Within the ever increasing list of such systems, mechanical metamaterials form a particularly prominent class due to the high contrast between the simplicity of the interactions between constituting elements, and the richness of the emergent physics \cite{Kadic2013,Christensen2015,Bertoldi2017NatureReview}. 

\begin{figure}
    \centering
    \includegraphics[width=0.97\linewidth]{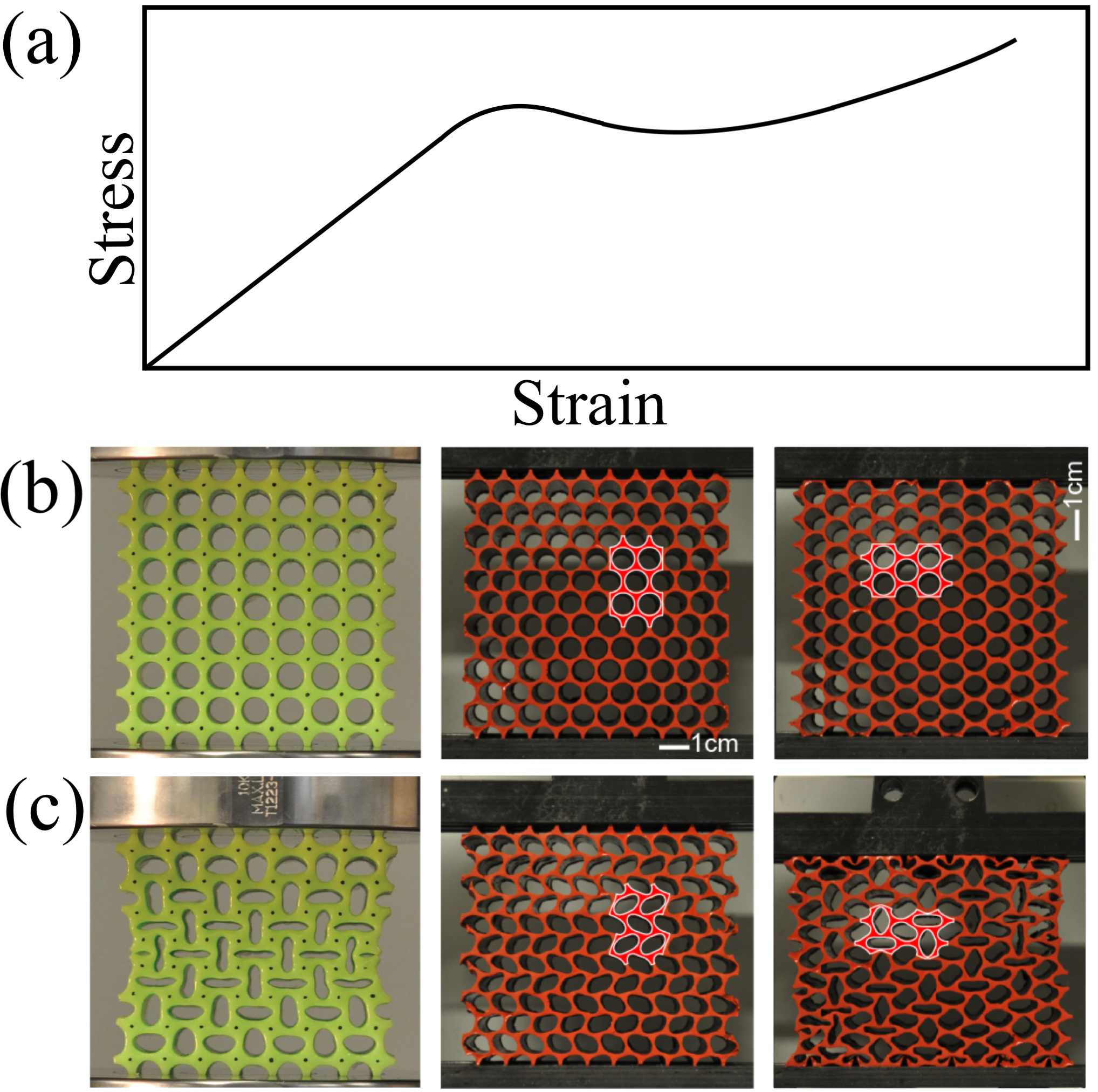}
    \caption{(a) A sketch of a typical stress strain curve for periodically perforated elastic metamaterial. (b,c) Square, triangular, and rotated triangular arrays of holes in an elastomer forming a mechanical metamaterial in an undeformed (b) and deformed (c) configurations. When subjected to uniaxial displacement from its boundaries a linear response is followed by an elastic instability. In the square lattice the instability is reflected via an alternate pattern of horizontal and vertical hole shapes whereas in the triangular lattice, due to frustration, the unstable mode forms either a zig-zag or a rosetta pattern. }
    \label{fig:Figure1-1}
\end{figure}

While the initial efforts focused on the design of mechanical metamaterials with unusual mechanical properties in the linear regime \cite{Kadic2013,Christensen2015}, more recently it has been shown that by embracing large deformations and instabilities these systems can achieve exotic functionalities \cite{Bertoldi2017NatureReview}. A prominent example of such nonlinear mechanical metamaterials consists of soft elastomeric matrix with embedded periodic array of holes  \cite{Mullin2007Pattern}. A typical stress-strain curve for such metamaterials is shown in \figref{fig:Figure1-1}(a). Under uniaxial compression, the linear response of the solid (at small loads) is a uniform deformation of the circular holes into ellipses, with their major axes oriented perpendicular to the external field. At higher loads, the system develops an instability and the stress plateaus. In the square lattice such instability results in the formation of a checkerboard pattern with the elongated holes take alternate horizontal and vertical orientations, whereas in the triangular lattices leads to  either a ``zig-zag'' or a rosetta pattern (see \figref{fig:Figure1-1}(c)), depending on the direction of the load. This \emph{spontaneous breaking of symmetry} is a telltale sign of an underlying nonlinear mechanism responsible for an instability~\cite{Bertoldi2010Negative}. Interestingly, this response is largely material independent, not only qualitatively but also quantitatively (e.g.~the critical strain at instability), implying a universal origin of the nonlinear mechanism. A central question then is about the emergent mechanics characteristic to such soft perforated elastic metamaterials out of an underlying  nonlinear theory of elasticity.

A theoretical analysis of the elastic problem requires solving the nonlinear equations of elasticity while satisfying the multiple free boundary conditions on the holes edges - a seemingly hopeless task from an analytic perspective. 
However, direct solutions of the fully nonlinear elastic equations are accessible using finite-element models, which accurately reproduce the deformation fields, the critical strain, and the effective elastic coefficients etc.~\cite{Bertoldi2010Negative}. The success of FE simulations in predicting the mechanics of perforated soft elastic materials confirms that nonlinear elasticity theory is a valid description, but emphasizes the lack of insightful analytical solutions to the problem.

A first attempt toward a theoretical explanation to this phenomenon was taken by Matsumoto and Kamien, who studied the interactions between holes based on the linear theory of elasticity \cite{Matsumoto2009Elastic, Matsumoto2012Patterns}. In their works they showed that the buckled patterns are consistent with energy minimizing configurations of interacting holes, if each hole is modeled as a pair of dislocations. While their work successfully captures the buckled modes, this approach is qualitative and cannot predict neither the critical strain at instability, nor the pre-instability linear response and the effects of holes on it. However, as a theory limited to describing the buckled state, Matsumoto \& Kamien's success implies that the concept of interacting holes can form the basis for an effective ``lattice'' theory of elastic metamaterials with periodic arrays of holes.

In this work we derive a new formalism that bridges the gap between the successful ``microscopic theory'' (nonlinear elasticity), and the macroscopic effective theory. As we will show, this formalism provides an insightful and intuitive description of perforated soft metamaterials without losing the quantitative capabilities of the microscopic theory. While the algebra might be somewhat technical, the qualitative picture that emerges from it is clean and elegant. Therefore, we structure the paper as follows: In Section~\ref{sec:qualitative} we qualitatively derive the main results of our analysis, using an analogy to a well known problem in electrostatics. In Section \ref{sec:themodel} we describe the full formalism and in Section~\ref{sec:results} we quantitatively compare its predictions to full numerical calculations. Finally, in Section~\ref{sec:discussion} we discuss possible promising directions.

\section{Qualitative picture}
\label{sec:qualitative}
There are two major challenges in writing an analytical theory: the multiple boundary conditions imposed by the holes, and the nonlinearity. As shown below, both these challenges can be tackled with the language of singular elastic charges. 
In what follows we will demonstrate that the phenomena can be approximately, but quantitatively, described in terms of interacting elastic charges with quadrupolar symmetry, located at the center of each hole. These are image charges, much like the image charges that are used to solve simple electrostatic problems. When the loading is weak (linear response), the interaction of the charges with the external field dominates and the quadrupoles align perpendicularly to the imposed stress. At higher stresses, due to geometrical non-linearities, the interaction between charges dominates their interaction with the external field, leading to the buckling instability that creates the patterns shown in Fig.~\ref{fig:Figure1-1}.  

\subsection{Electrostatic analogy}
To see how this comes about, it will be useful to recall a familiar pedagogical problem in 2D electrostatics that will serve as an analogy for the corresponding problem in elasticity. Consider a circular conductive shell in the presence of a uniform external electric field. Solving for the resultant field requires a solution of Laplace's equation with specific boundary conditions on the conductive surface.
One particularly insightful method to solve this equation, introduced in elementary physics classes, is the method of image charges. The trick is that placing ``imaginary'' charges outside the domain of interest (i.e.~inside the shell) solves by construction the bulk equation, and wisely chosen charges can also satisfy the boundary conditions. Indeed, the problem is solved exactly by placing a pure dipole at the shell center. From the perspective of an observer outside the shell, the presence of the conductive surface is indistinguishable from that of a pure dipole. Thus, the concept of image charge not only opens an analytic pathway for solving the problem, but also provides intuition about the solution, and specifically on the physical effect of boundaries.

We note two properties of the solution which will have exact analogs in elasticity: first, the imaginary charge is a dipole, not a monopole. Electrostatic monopole image charges are disallowed because they are \textit{topologically conserved}. That is, the net charge in a given region can be completely determined by a surface integral on the region's boundary (Gauss' theorem). Second, the magnitude of the dipole moment turns out to be proportional to the external field and to the circle's area (in 2D).

How do we find the correct image charges?
A common strategy is to find them by enforcing the boundary conditions. This works only in cases where the image charges can exactly solve the problem. An alternative approach is via energy minimization, which gives an approximate solution when the exact one cannot be represented by a finite number of image charges. In fact, a potential $\phi$ that satisfies the equation and boundary conditions is a minimizer of the energy
\begin{equation}
    F = \int_\Omega  \tfrac{1}{2}|\vec{\nabla} \phi-
    \bm{E}^{ext}|^2 \dif S - \oint_{\partial \Omega} \rho \, \phi  \, \dif l
    \label{eq:ElecFE}
\end{equation}
where $\bm{E}^{ext}$ is the imposed external field, $\Omega$ is the problem domain (e.g. $\mathbb{R}^2$ with a circle taken out) and $\partial\Omega$ is its boundary.  The function $\rho$ is a Lagrange multiplier enforcing a constant potential on the conducting boundary,
see \appref{app:VariationalApproach}.
For the problem described above, after guessing a solution in the form of a single dipole, its magnitude can be found by minimizing the energy \eqref{eq:ElecFE} with respect to the dipole charge and $\rho$. The result satisfies the boundary conditions exactly.

Consider now a harder problem: an array of conducting circular shells in an external electric field, introducing the complication of multiple boundary conditions.
In contrast with the single shell problem, guessing a finite number of image charges that will balance boundary conditions is impossible: The image charges are now reflections of the external field, but also of all other image charges. Therefore, in general, the image charge in each shell is composed of infinite number of multipoles.
While an exact solution is hard to guess, by minimizing the energy we can nonetheless obtain an approximate solution. Each circular shell is going to be polarized, and the dominant image charge inside each shell is dipolar: higher order multipoles decay faster in space and therefore would have smaller energetic contributions. Thus, guessing a solution in terms of dipoles is a controlled approximation and accounting for higher order multipoles inside each shell would improve the accuracy of the solution.
Specifically, we guess an ansatz of the form
\begin{equation}
    \phi\brk{\xvec} = \sum_i \mathbf{p}_i \cdot \pmb{\phi}_{\mathbf{p}} \brk{\xvec - \xvec_i}
    \label{eq:EAsubs}
\end{equation}
where $\mathbf{p}_i$ is the image dipole vector located at $\xvec_i$ (the center of the $i$-th conducting shell), and $\pmb{\phi}_{\mathbf{p}}$ is the well known solution for the potential of a single electric dipole. The energy can be written as a quadratic form in the unknown charges $\mathbf{p}_i$:
\begin{equation}
    F =  \mathbf{M}_{ij} \mathbf{p}_i \mathbf{p}_j  -  \mathbf{m}_i \mathbf{p}_i,
    \label{eq:LinFE}
\end{equation}
where
\begin{equation}
\begin{split}
    \mathbf{M}_{ij} &= \frac{\epsilon_0}{2} \int_\Omega \brk{\vec{\nabla} \pmb{\phi}_{\mathbf{p}} (\xvec - \xvec_i)} \brk{\vec{\nabla} \pmb{\phi}_{\mathbf{p}}(\xvec - \xvec_j)} \dif S\ , \\ 
    \mathbf{m}_i &= \oint_{\partial \Omega}  \pmb{\phi}_{\mathbf{p}} (\xvec-\xvec_i) \rho_i\brk{\xvec} \dif l\ .
\end{split}
\label{eq:MInt}
\end{equation}
The matrix $\mathbf M$ quantifies interactions between image dipoles in different shells, and $\mathbf m$ quantifies interactions of these dipoles with the external field. Since the potential of a dipole is known in explicit analytical form, calculating $\mathbf M$ and $\mathbf m$ is a trivial task of integration\footnote{One should also decompose $\rho$ in terms of the dipolar fields, but since this is just an analogy, we do not go into these details here.}. Then, minimizing the energy \eqref{eq:LinFE} is straightforward. 

\subsection{The elastic problem}
All the above concepts can be translated, with some modifications, to elasticity theory. The  linear elastic analog of the single conducting shell problem happens to be a famous example, solved by Inglis in 1913~\cite{Inglis1913}: a circular cavity in an infinite 2D elastic medium, subject to remote stress. Mathematically, the problem amounts to solving the bi-harmonic equation for the Airy stress function and, like the electrostatic analog, the Inglis solution is equivalent to a pure imaginary \emph{elastic charge} at the shell center \cite{Moshe2019PRL, Moshe2019PRE}. The charge is a quadrupole and in the linear theory its magnitude is proportional to the applied stress and to the hole's area (in 2D). But what are elastic charges?

A geometric approach to elasticity uncovers the mathematical nature of elastic charges. The physical quantity associated with elastic charges is Gaussian curvature, that is, a monopolar charge is a singular distribution of Gaussian curvature. As an example, consider a thin conical surface confined to the flat euclidean plane. The stressed state of the flattened cone reflects a geometric incompatibility between the flat embedding space and the conical reference state. The incompatibility is quantified by the Gaussian curvature of the reference state, which in the case of a cone is a delta-function singularity at the apex \cite{Seung88, Moshe2015PNAS}. 

Since the Gaussian curvature of the reference state acts as a singular source of elastic fields, it can be interpreted as an elastic charge. In crystalline materials, the monopole singularity described above is manifested as a disclination \cite{Seung88,Moshe2015PNAS}. A dipole of elastic charges, i.e.~a pair of disclinations of equal and opposite magnitude, forms a dislocation \cite{Seung88}. Lastly, a quadruplolar charge, like the one which solves the circular hole problem, is realized as a dislocation pair with equal and opposite Burgers vectors, also known as the Stone-Wales defect in hexagonal lattices. In the context of continuum theory the elastic quadrupole is known as an elliptic Eshelby inclusion, i.e.~an irreversible deformation of a circular domain into an ellipse~\cite{Eshelby1957, Dasgupta2012}. Another realization of a quadrupole is a force-dipole applied locally to an elastic substrate, e.g. by adherent contractile biological cells~\cite{Safran2013}. 

Like in the electrostatic case, the fact that the lowest order multipole that solves the hole problem is a quadrupole is a direct consequence of a conservation theorem. In electrostatics, local creation of monopoles is disallowed by conservation. In elasticity, both the monopole (Frank's vector) and the dipole (Burgers' vector) are conserved \cite{LandauElasticityBook}. For a rigorous derivation of all these results, see \cite{Kupferman2013ARMA}.

With the method of image charges on one hand, and the concept of elastic charges on the other hand, we can now attack the elastic problem of soft metamaterials containing array of holes.
This problem can be solved by placing imaginary charges in the center of each hole, but these charges also create their own image charges inside other holes, like in the electrostatic case of an array of conducting shells. That is, the complex interactions between holes can be described in terms of multiple image charges interacting with each other and with the imposed external field. 
As in the electrostatic case, an approximate solution for a given external load can be derived by guessing a solution for which the elastic fields are dominated by the lowest order non-topological charges, that is, imaginary quadrupoles. 

\begin{figure}
    \centering
    \includegraphics[width=1\linewidth]{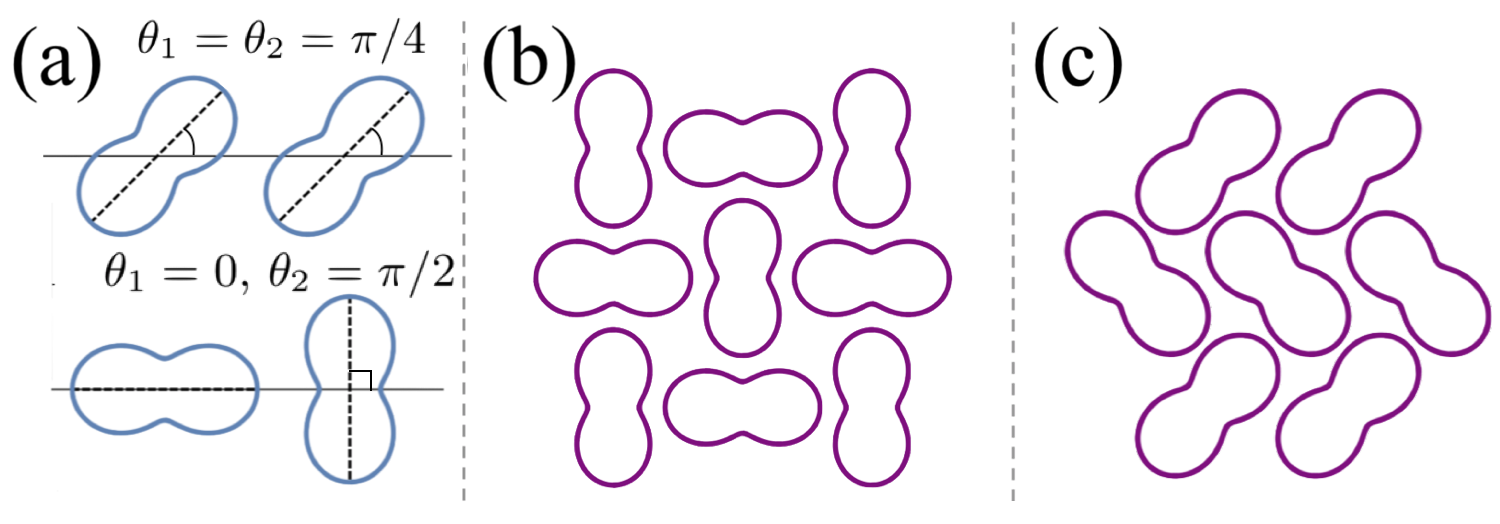}
    \caption{Interacting elastic quadrupoles, illustrated by the deformation fields they induce on the holes edges. (a) Two energy minimizing configurations of quadrupoles of fixed magnitudes and free orientations. In the top configuration we have $\theta_1=\theta_2=\pi/4$ while in the bottom one $\theta_1=0$, $\theta_2=\pi/2$. (b) An array of quadrupoles on a square lattice minimizing their interaction energy with nearest and next to nearest neighbors, as given by equation \eqref{eq:QuadInt}. The relative orientation of any nearest-neighbor pair is like the bottom row of panel (a), and that of next-nearest pairs is like the top row. (c) Like panel (b), but for a triangular lattice. 
    }
    \label{fig:Figure2}
\end{figure}

\subsection{Interacting quadrupoles}
Let's assume for the moment that the solution is indeed composed of a quadrupole located at the center of each hole. As a first attempt, let us also assume that the  magnitude of all quadrupole is fixed and they are free to rotate (this assumption is in fact holds in the square and straight triangular lattice shown in \figref{fig:Figure1-1}). This picture, of interacting rotating quadrupoles is very close in spirit to the phenomenological description of Matsumoto \& Kamien's~\cite{Matsumoto2009Elastic}, who described each hole as a pair of opposite dislocations, i.e.~an elastic quadrupole. To proceed, we need to understand the interaction between two pure elastic quadrupoles in an infinite elastic medium. For two quadrupoles of magnitude $Q_1, Q_2$ and orientations $\theta_1, \theta_2$ ($\theta_i$ is measured with respect to the line connecting the two quadrupoles, see Fig.~\ref{fig:Figure2}a), the interaction energy is~\cite{Moshe2015PRE}
\begin{equation}
E = \frac{Q_1 Q_2}{\pi r} \cos\brk{2\theta_1 + 2\theta_2}.
\label{eq:QuadInt}
\end{equation}
This energy is minimized for configurations satisfying $\theta_1 + \theta_2 = \pi/2$, which is a 1 dimensional continuum of minimizers. \figref{fig:Figure2}(a) presents two such optimal configurations.

What is the optimal configuration of a lattice of quadrupoles? For a square lattice, if only nearest neighbors interactions are taken into account, two distinct energy minimizing configurations satisfy the condition $\theta_1 + \theta_2 = \pi/2$ for all neighboring quadrupoles. (i) All quadrupoles having an angle of $\pi/4$ relative to the horizontal axis, as in the top row of Fig.~\ref{fig:Figure2}a (ii) A checkerboard pattern of horizontal and vertical quadrupoles (as in the bottom row of Fig.~\ref{fig:Figure2}a and Fig.~\ref{fig:Figure2}(b)). However, the checkerboard pattern has a lower energy because it also minimizes the interaction between quadrupoles on opposing sides of the unit square diagonal, i.e.~next-nearest-neighbors. Note that this is exactly the pattern of the buckled state of the square lattice, cf.~\figref{fig:Figure1-1}.

Unlike the square lattice, the symmetries of the triangular lattice are incompatible with those of the interacting quadrupoles. This incompatibility is manifested through a non-zero ground-state energy. Direct minimization of nearest-neighbors interactions energy with respect to quadrupoles orientations we find the resulting pattern shown in \figref{fig:Figure2}(c).
As before, quadrupoles orientations are in agreement with the observed unstable mode. 
The Rosetta pattern observed in the rotated triangular lattice shown in \figref{fig:Figure1-1}(c), however, is not captured by this simplified model, since in it the magnitudes of the quadrupoles are not uniform. 


\subsection{Collecting the pieces}
The conclusion from the previous section is that the unstable modes resemble a collection of interacting quadrupoles.
We suggest that rigorously describing the system as a collection of interacting quadrupoles is a perturbative approximation of the  full solution:
At low stresses, all quadrupoles are aligned with the external field.
At higher stresses the metamaterial buckles and, as we just seen, the buckled states are consistent with a model of interacting quadrupoles. This suggests that the post-instability response is dominated by charge-charge interaction rather than interactions of charges with the external load.

We emphasize, however, that this picture does not have a (linear) electrostatic analog. In linear systems, the induced charges are always proportional to the external loading ($E^{ext}$ in \eqref{eq:ElecFE}) and therefore the interaction between themselves cannot, by construction, dominate their interaction with the external field. The mechanism described above is manifestly nonlinear and requires a generalization of the electrostatic arguments. The observed instability emerges from a universal nonlinearity, which is inherent to elasticity, and does not have an electrostatic analog. 
Below we show how the framework of interacting charges can be expanded to account for all these effects.

\section{The method}
\label{sec:themodel} 
The fundamental field in the theory of elasticity is the displacement field $\d$, which measures the spatial movement of material elements from a reference position to its current one. Local length deformations are quantified by the strain tensor $\us$~\cite{landau},
\begin{equation}
\us = \frac{1}{2} \brk{\nabla \d + \nabla \d^T + \nabla \d^T \cdot \nabla \d}.
\label{eq:Strain}
\end{equation}

The elastic energy density, which results from local length changes, can be written as a function of $\us$. Linear elasticity is a leading-order perturbation theory for small deformations and therefore $E$ is written as a quadratic function of $\us$, also known as Hookean energy
\begin{equation}
E = \left\langle \us ,\us \right\rangle +\mathcal{O}(\us^3)\ .
\label{eq:en}
\end{equation}
Here $\left\langle \mathbf{v},\mathbf{u} \right\rangle \equiv \int_{\Omega} \frac{1}{2}\mathbf{v} \A \mathbf{u} \, \dif S$, is an integration over the domain $\Omega$ of the contraction of the tensor fields $\mathbf{u}, \mathbf{v}$ with a 4-rank tensor $\A$, known as the elastic (or stiffness) tensor, which encodes material properties such as Young's modulus and Poisson's ratio.

Although the energy is quadratic, the theory as presented above is still nonlinear due to the $\nabla \d^T \cdot \nabla \d$ term in the strain \eqref{eq:Strain}.  Negelcting it (assuming $\nabla \d\ll 1$) yields the familiar theory of linear elasticity~\cite{LandauElasticityBook}. That is, linear elasticity is obtained by performing two conceptually distinct linearizations: a \emph{rheological} linearization, neglecting higher-order material properties (the $\mathcal{O}(\us^3)$ term in \eqref{eq:en}), and a \emph{geometrical} linearization, neglecting the quadratic term in \eqref{eq:Strain}. 
In the former, the neglected nonlinear behavior is rheological and therefore material-specific. In the latter, the neglected terms are geometrically universal  and relate to rotational invariance. Since, as described above, the nonlinear mechanics of elastic metamaterials with arrays of holes is largely material independent, it is reasonable to speculate that a suitable analytical description of the system is that of a nonlinear geometry with a quadratic (Hookean) energy.  Therefore, we take Eqs.~(\ref{eq:Strain}-\ref{eq:en}) to be the governing equations in this work.

Numerical analysis has confirmed the applicability of these equations in two respects: first, a full numerical solution of the governing equations accurately reproduces experimental results~\cite{Bertoldi2010Negative}. Second, calculations show that even in buckled state, which is clearly a nonlinear response, $\nabla \d$ is of order unity due to almost-rigid rotations of the junctions between holes, invalidating the geometric linearization. However, the non-linear strain (Eq.~\eqref{eq:Strain}) is small due to cancellation of the linear and quadratic terms, justifying the rheological linearization in Eq.~\eqref{eq:en}. In the remaining of this paper we put the method of image quadrupoles together with this elastic nonlinearity to construct an effective nonlinear theory. 

\subsection{Bulk energy}
We express the total deformation in the system as induced by quadrupoles sitting at the centers of holes, very similar to equation \eqref{eq:EAsubs}. Using a recent generalization of the method of Airy stress function, which allows solving elastic problems with arbitrary constitutive relations, strain definitions or reference states \cite{Moshe2014ISF,Moshe2015PRE}, we perform a perturbative expansion of the nonlinear quadrupolar fields~\footnote{In fact, a careful analysis of the elastic equations reveals that there are two distinct types of elastic monopoles, and consequently also two types of quadrupoles. For succinctness in the text we refer to quadrupoles in a general manner, but in the actual calculations we do take into account both types of quadrupoles in each hole. A detailed calculation is presented in  \appref{app:Hex}.}. Symbolically, the displacement induced by a single charge $q_{\alpha\beta}$ located at $\xvec _i$ is written as
\begin{equation}
    \d(\xvec) = \sum_i\sum_{\alpha\beta} q_{\alpha\beta i}\, \d^{(1)}_{\alpha\beta  }(\xvec-\xvec_i) + 
	{q_{\alpha\beta i}}^2\, \d^{(2)}_{\alpha\beta}(\xvec-\xvec_i) + \cdots
    \label{eq:DispExpansion}
\end{equation}
where $\d^{(n)}_{\alpha\beta}$ is the displacement to $n$-th order associated with the charge $q_{\alpha\beta i}$. For the first time we solve the geometrically nonlinear fields associated with elastic multipolar charges, and give the detailed derivation in \appref{app:NLSF}. In addition to image charges at the hole centers, we also allow uniform elastic fields, which within the formalism are described as quadrupolar charges located at infinity. 

For notational simplicity, it is easier to denote the collection of all components of all charges, either at hole centers or at infinity, by a single vector $\mathbf Q$, replacing the 3 indices $\alpha, \beta, i$ by a single index. Combining the ansatz \eqref{eq:DispExpansion} with the elastic energy Eqs.~\eqref{eq:Strain}-\eqref{eq:en} we obtain
\begin{align}
E &= \sum_{ij}\, \tfrac{1}{2}\M^{(1)}_{ij}\, Q_i \, Q_j  + \frac{1}{6}\M^{(2)}_{ijk}\, Q_i \, Q_j \, Q_k  +\dots \ ,
\label{eq:QuadEnergy}
\end{align} 
where
\begin{align}
\begin{split}
\M^{(1)}_{ij} &= \left\langle \us^{(1)}_{i}, \us^{(1)}_{j}\right\rangle\\
\M^{(2)}_{ijk} &= \left< \us^{(1)}_i, \us^{(2)}_j \right> \delta_{jk} +\left< \us^{(2)}_i, \us^{(1)}_j \right> \delta_{ik}\ .
\end{split}
\end{align}
Here, $\us ^{(k)}_j$ is the strain field derived from the displacement field $\mathbf{d}^{(k)}_j$ and $\delta_{ij}$ is the Kronecker delta.

The matrix $\M^{(1)}$, similar to the electrostatic alanog $\mathbf M$ of Eq.~\eqref{eq:MInt}, has a simple interpretation: it is a positive-definite matrix that quantifies pair-interactions between charges, taking into account their relative position and the geometry of the domain. Similarly, $\M^{(2)}$ describes the interactions between triplets of charges, and so on.

\subsection{External loading}
In the electrostatic example above we dealt with infinite systems where the external loading was imposed by a bulk energetic term ($\bm{E}^{ext}$ in Eq.~\eqref{eq:ElecFE}). It is possible to include such a term in the elastic theory too, but in this work we want to analyze the case most commonly encountered in reality: a finite system with displacement-controlled boundary conditions, as in Fig.~\ref{fig:Figure1-1}. This requires a different approach and there are a few ways in which these boundary conditions can be introduced within our formalism. We found that, in the context of the lattice-hole geometry, treating them as constraints on the unknown charges $\mathbf Q$ is the most convenient approach. 

As discussed above, the boundary conditions cannot be satisfied exactly when expressing the relevant fields with a finite number of charges. However, an approximate solution can be obtained by demanding that the boundary conditions will be satisfied \textit{on average} in a particular region. Consider the geometry of the system, depicted in Fig.~\ref{fig:Figure1-1}: the top and bottom boundaries of the lattice are loaded by a rigid plate. The actual contact points between the system and the loading mechanism are a discrete set of ligaments. Focusing on one of them, the average displacement on the boundary is given
\begin{align}
\bar{\d} = \sum_i \N_i^{(1)}Q_i + \N_i^{(2)}{Q_i}^2 + \cdots\ ,
\end{align}
where $\N^{(i)}$ can be expressed by explicit integration of Eq.~\eqref{eq:DispExpansion} over the ligament. Imposing a given average displacement on a set of ligaments translates to a collection of nonlinear constraints on the charges, one for each ligament. That is, the constraints on the charges are
\begin{align}
 \left(\sum_{i} Q_i\N_{ij}^{(1)}+ {Q_j}^2 \N_{ij}^{(2)} + \cdots\ \right)-\d^{ext}_j=0,
 \label{eq:constraints}
\end{align}
where $\d^{ext}_j$ is the imposed displacement on the $j$-th ligament and $\N^{(i)}$ is a $N\times c$ matrix. Here, $c$ is the number of constraints and $N$ is the number of charge components, i.e.~the length of the vector $\mathbf Q$. 

In this formalism, finding the charges that best approximate the boundary conditions amounts to minimizing the non-linear energy \eqref{eq:QuadEnergy} under the nonlinear constraints of Eq.~\eqref{eq:constraints}.
\begin{figure}
    \centering
    \includegraphics[width=1\linewidth]{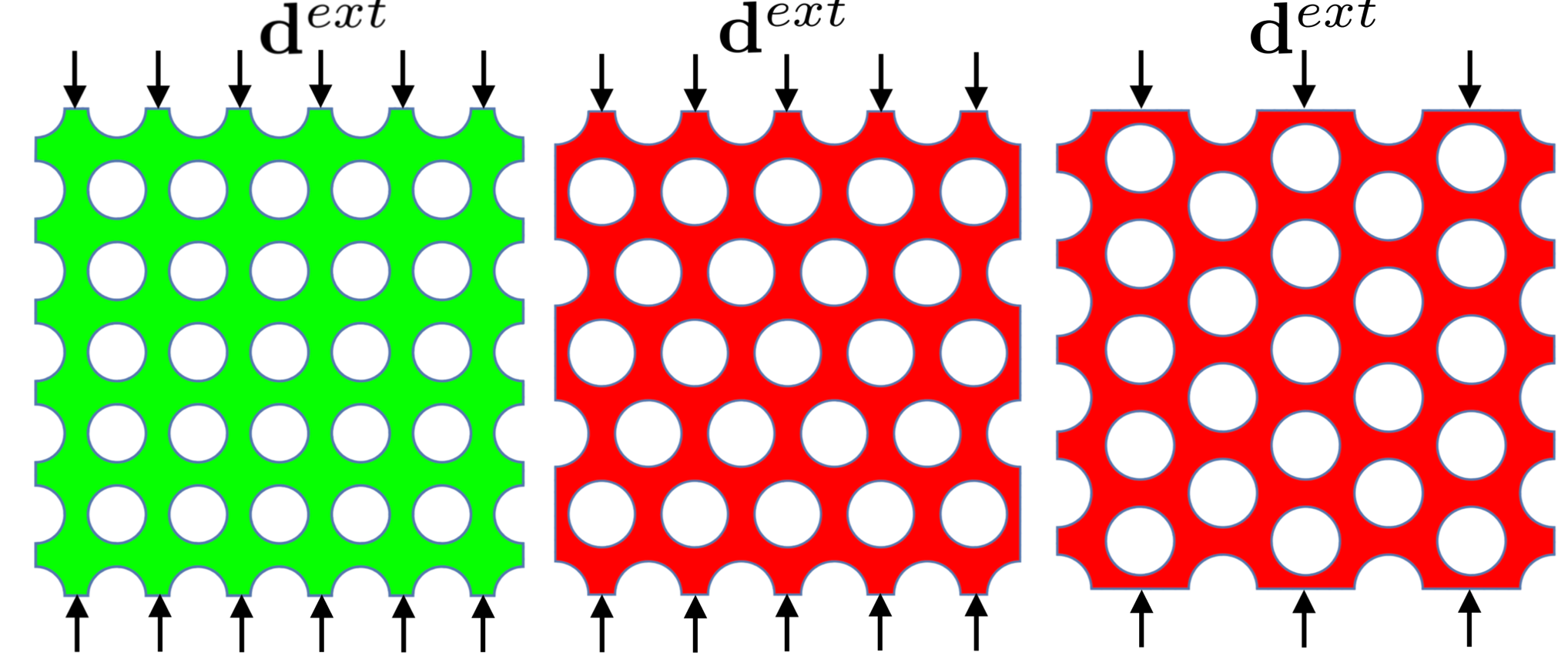}
    \caption{The three prototypical lattices studied in this work. A qquare, triangular, and rotated triangular lattices of circular holes with uniform size subjected to external displacement $\mathbf{d}^\text{ext}$ applied on the ligaments forming the boundaries.
    }
    \label{fig:System}
\end{figure}

\section{Results}
\label{sec:results}

Here we implement the method of image quadrupoles to three prototypical lattices: A square lattice, a triangular lattice, and a triangular lattice rotated by $\pi/2$, as shown in \figref{fig:System}. In our analysis the lattices contains approximately $9\times9$ holes, and are characterized by their porosity, defined as the fractional area of holes. Here we analyze systems with porosity that ranges from $p=0.3$ to $p=0.7$ (the percolation limit is at $p_c=0.78$).
To test the theory we compare the results with direct numeric simulations of the full equations, which is known to agree very well with experiments~\cite{Bertoldi2010Negative}.

\begin{figure}
	\centering
	\includegraphics[width=\columnwidth]{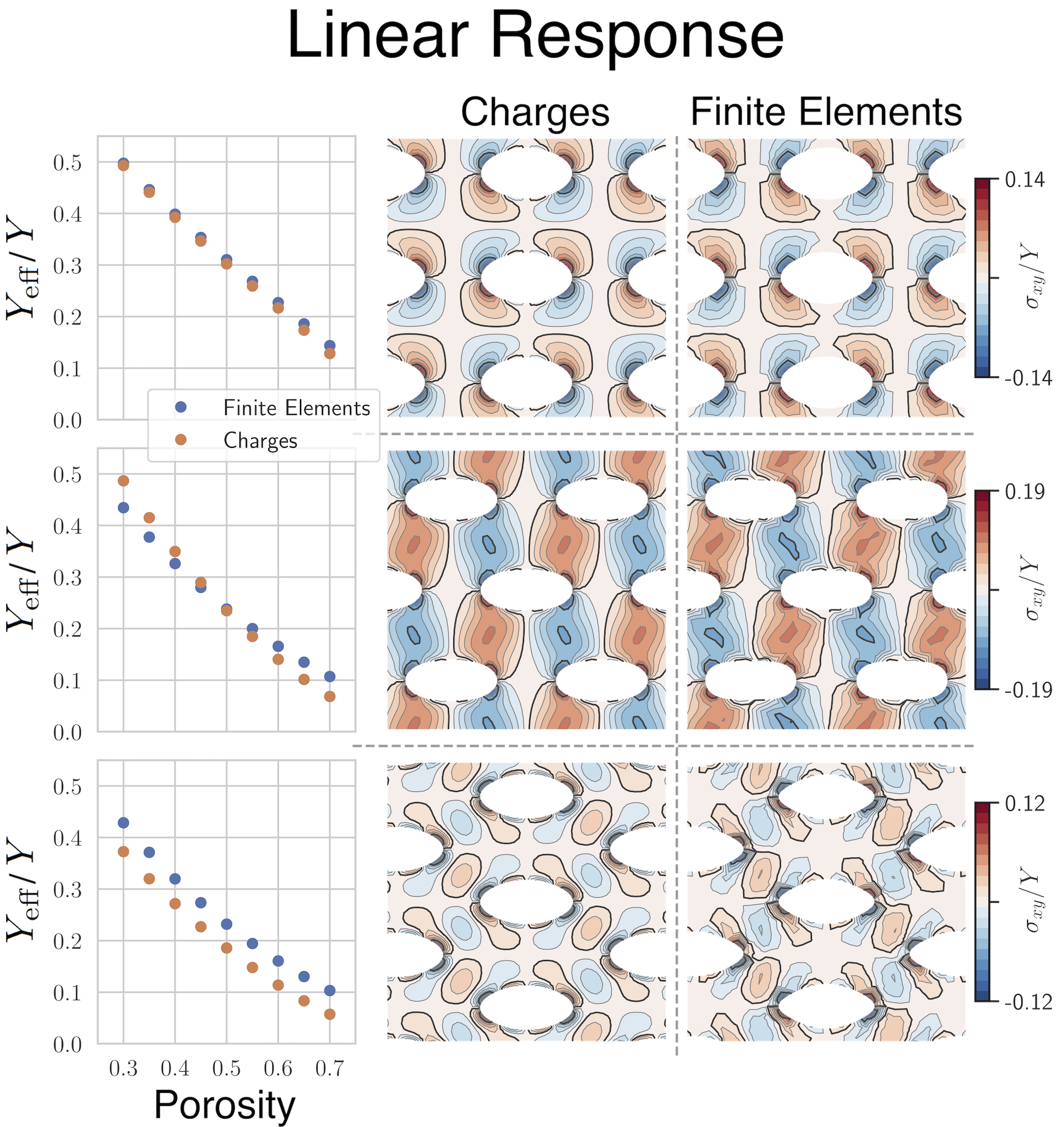}
	\caption{Comparison between the elastic-charges calculation and a direct fully nonlinear numerical solution in the linear regimes for three different lattices. The left column shows the effective Young's modulus as function of porosity (blue for finite-element, orange for elastic-charges). Each point represent the slope in the stress-strain curve of a system with the corresponding hole pattern and porosity. Center and right panels show representative fields of the  $\sigma^{xy}$ component of the stress field distribution plotted on top of the strained configurations with porosity $p=0.3$. 
	}
	\label{fig:LinRes}
\end{figure}

\begin{figure*}
	\centering
	\includegraphics[width=0.75\linewidth]{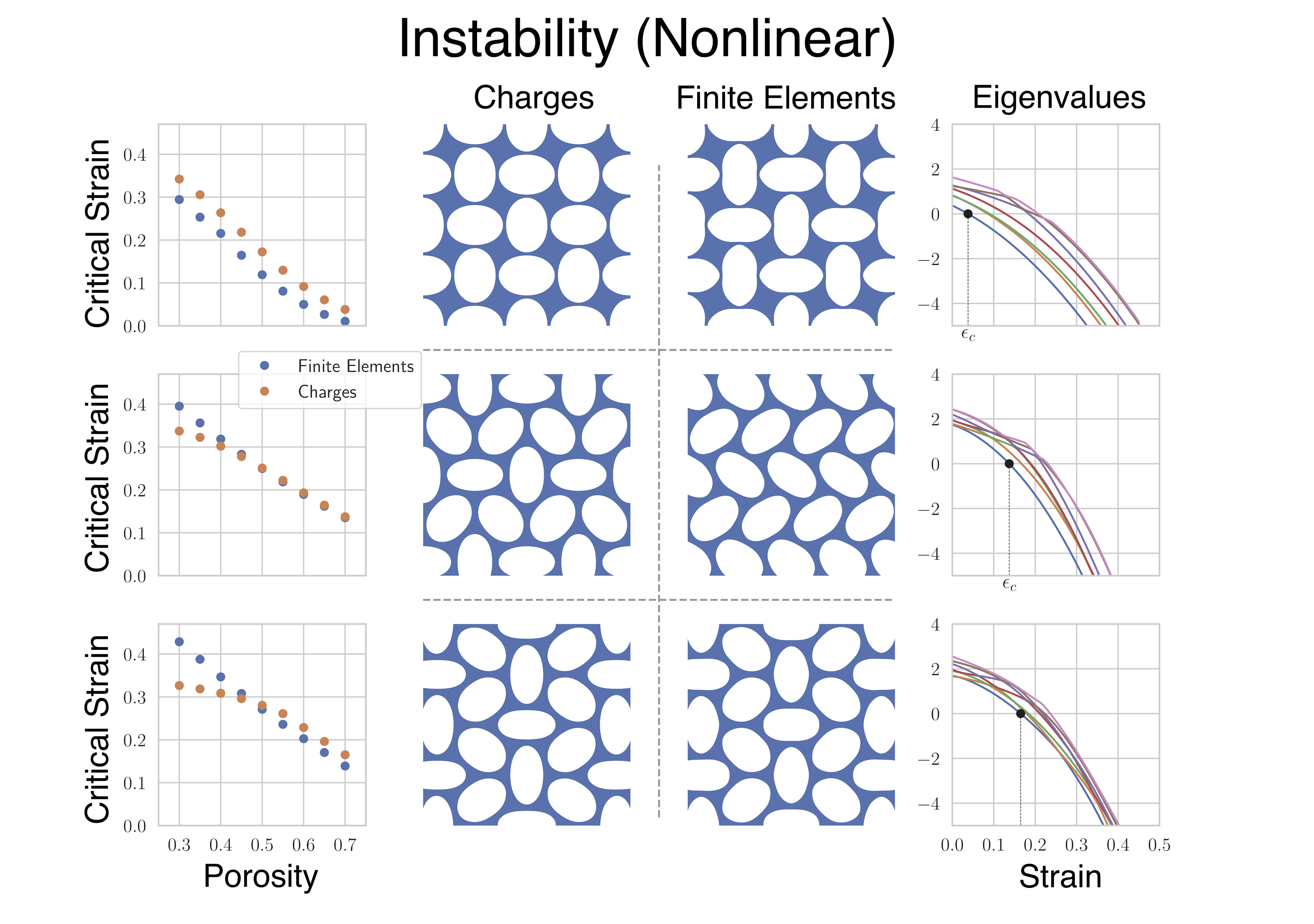}
	\caption{Comparison between the elastic-charges calculation and a direct fully nonlinear numerical solution at the onset of instability for three different lattices. In the left panel we plot the critical strain as function of porosity  (blue for finite-element, orange for elastic-charges). Center panels show representative unstable modes from the elastic-charges approach, and holes configurations beyond the instability from numerical simulations, both with porosity $p=0.7$. In the right panel we plot the eigenvalues as function of strain, demonstrating the formation of instability and the congestion of vanishing eigenvalues at the onset of instability.
	}
	\label{fig:NonLinRes}
\end{figure*}

\subsection{Linear response}
We begin by analyzing the linear response of the system under small displacements. In this limit, only the leading order contributions are considered. That is, we minimize the quadratic energy $E= \sum_{ij}\, \tfrac{1}{2}\M^{(1)}_{ij}\, Q_i \, Q_j$ under the set of linear constraints $\sum_j\N_{ij}Q_j= \d^{ext}_i$. This is a trivial exercise in linear algebra, and the desired charges are given by
\begin{align}
\mathbf{Q}^* = \M^{-1}\N\left(\N^T\M^{-1}\N\right)^{-1}\d^{ext} \ .
\label{eq:Q}
\end{align}

With $Q^*$, the solution can be written in terms of Eq.~\eqref{eq:DispExpansion} and any property of interest can be extracted. For example, in left column in Fig.~\ref{fig:LinRes} we plot the coarse-grained Young modulus $Y_\text{eff}$ as function of porosity, measuring the system's compliance for uniaxial loads, i.e.~its effective spring constant. It is defined by the ratio of the average compressive stress to the compressive strain. Comparison to direct numerical simulations shows that the formalism quantitatively captures the coarse-grained response of the system. 
In addition, in middle and right columns in \figref{fig:LinRes} we plot the spatial distribution of the stress field $\sigma^{xy}$ for a representative porosity and imposed strain, plotted on top of the deformed configurations, showing a favorable agreement also in the detailed spatial structure of the solution. 
We emphasize that the charge formalism has no free parameters to fit. 

A slight difference in the deformation field is observed in the case of rotated triangular lattice, as shown at the bottom of middle and right columns. 
This difference reflects the fact that using only quadrupolar charges cannot fully describe the solution. To capture these details, higher order multipole are needed. 


\subsection{Instability (nonlinear response)}
Encouraged by the success of image charge method in the linear regime, we now proceed to studying the nonlinear instability of the system. In particular, we are interested in the critical strain at the onset of instability, and the unstable modes.
 

The stability of the system is determined by the Hessian of the energy which in the linear response regime is simply $2\M^{(1)}$. It is guaranteed to be positive definite and the system is thus stable. Expanding to the next order in $\d^{ext}$, we find that the Hessian reads
\begin{align}
\label{eq:NLHessian}
H_{ij} &=
2\M^{(1)}_{ij}+\\
\nonumber
&4\left[Q^*_i \M^{(2)}_{ij}+Q^*_j \M^{(2)}_{ji}+\sum_k \M^{(2)}_{ik} Q^{*}_k \delta_{ij}\right]
\end{align}
where no summation is intended on $i$ and $j$. In addition, the displacement constraints of Eq.~\eqref{eq:constraints} should also be corrected to next-leading order. This technical calculation is done in detail in \appref{app:NLImplementation}. 

The charges in the linear solution, $\mathbf{Q}^*$, are proportional to the imposed displacement $\d^{ext}$, cf.~Eq.~\eqref{eq:Q}. This means that the correction to the Hessian (the bracketed term in Eq.~\eqref{eq:NLHessian}), as well as the correction to the displacement constraints, is also linear in $\d^{ext}$. When the imposed displacement is large enough, the constrained Hessian can become singular, i.e.~one of its eigenvalues can vanish. This is the onset of instability.

We note that this calculation is in line with the intuitive picture described above: For small loads (i.e.~in the linear regime) the dominant interaction is that of the charges with the external loading and with themselves, quantified respectively by $\N^{(1)}$ and $\M^{(1)}$. In this regime the solution is linear in $\d^{ext}$ and given by Eq.~\eqref{eq:Q}. It is stable because $\M^{(1)}$ is positive definite. For larger loads, the interaction of the induced charges with themselves, quantified by $\M^{(2)}$ becomes important and eventually destabilizes the linear solution.

The left column of \figref{fig:NonLinRes} shows the critical strain, i.e.~the strain at which the Hessian becomes singular, as function of porosity for the three different lattices. Our method is in good quantitative agreement with the full numerical simulations, except possibly at very low porosities. This happens because smaller porosities lead to larger critical strains, making the image charges magnitudes larger. Because our method is a perturbative  expansion in the charge magnitude, its accuracy deteriorates when the charges are large. This effect is more noticeable in the triangular lattices (two bottom rows of \figref{fig:NonLinRes}).

For each lattice, we also plot the unstable eigenmode associated with the vanishing eigenvalue. Representative ones are plotted in the two center columns in \figref{fig:NonLinRes}. 
In two out of the three cases shown here, the unstable modes computed with our method agree with those found in finite element simulations. 
The case shown in the second row of \figref{fig:NonLinRes} is an exception. It might come as a surprise that the formalism properly identifies the critical strain, i.e.~the load where a specific eigenmode becomes unstable, while the mode itself is not the right one. A deeper investigation reveals that many eigenmodes become unstable almost simultaneously, making it difficult to pinpoint the least stable one. This is clearly seen in the right column in \figref{fig:NonLinRes}, where at the onset of instability many eigenvalues are densely distributed close to the vanishing one. The zigzag-like mode, like the one predicted by finite element simulations and by the interacting quadrupole model of \figref{fig:Figure2}, also becomes unstable at a similar strain.

\section{Summary and discussion}
\label{sec:discussion}
We introduced a formalism that identifies image elastic charges as the basic degrees of freedom of soft perforated elastic metamaterials. The continuum elastic problem, which contains multiple boundary conditions, is reduced to a simpler problem of interacting elastic quadrupoles.

A central advantage of the elastic-charges approach is its conceptual aspect, in that it offers understanding and intuition about holes patterns before making any calculation. Both the linear response pattern and the buckled state can be qualitatively understood easily, as well the instability mechanism.

In addition, we found very good quantitative agreement between the theory of elastic charges and detailed nonlinear finite-element analysis. This includes the effective Young's modulus, the stress field distribution, the critical loads at the onset of instability, the unstable modes, and more.

Lastly, the charge formalism is also beneficial from a computational perspective, since it vastly reduces the number of degrees of freedom in the problem. For a finite-element simulation to be reliable, the mesh must contain at least a few dozen points per hole. In the simulations reported in this work, a reasonable accuracy demanded around $~10^4$ mesh points. The elastic charge formalism, on the other hand, requires a handful of degrees of freedom per hole. In the calculations reported here, this number was usually around 300. The few dozen charge method calculations in this work run on a stand laptop within a matter of minutes, combined.

However, we emphasize that at its present form, the model cannot serve as an alternative to the detailed finite-element analysis. For example, while our theory correctly describes mechanical properties prior to, and at the onset of instability, it is not valid beyond the instability: Since our theory expands the energy only to 3rd order, the post-instability energy does not have a minimum. Analyzing the post-instability response requires going to the next order, with a quartic energy functional. This is a topic for future work.

Looking forward, we suggest that this approach might open the way for importing techniques and ideas from Statistical Mechanics to the study of perforated elastic metamaterials. For example, we are currently investigating the effect of structural disorder by introducing randomness to the mechanical interactions between the charges (i.e.~randomness in the interaction matrices $\M$ and $\N$). Another direction, for future work, would be coarse-graining the model to develop a field theory where the quadrupolariztion is a continuous field. This would be the analog of dielectric materials described by distributing induced electric dipoles, but with richer response.

\begin{acknowledgments}
	Y.B.S acknowledges support from the James S.~McDonnell post-doctoral fellowship for the study of complex systems.
	M.M. acknowledges useful discussion with David R. Nelson, Mark J. Bowick, and Eran Sharon.
	This research was supported by the Israel Science Foundation (grant No. 1441/19).
\end{acknowledgments}

 \appendix
 \section{Nonlinear incompatible stress function theory}
 \label{app:NLSF}

\paragraph*{The fields induced by elastic charges} 
Calculation of the nonlinear fields induced by elastic charges requires the solution of equilibrium equations, reflecting force balance on each material element
\begin{equation}
\text{Div} \, \bfsigma = 0.
\label{eq:div}
\end{equation}
Here $\bfsigma$ is the stress tensor, defined by variational derivative of the energy with respect to strain, leading to a locally linear stress-strain relation in Hookean elasticity 
\begin{equation}
\boldsymbol{\sigma} = \A \mathbf{u}
\label{eq:Hook}
\end{equation}
with $\A$ the elastic tensor, encoding local material properties.
The seemingly innocent equilibrium equation is in fact highly nonlinear, with the operator $\text{Div}$ being a generalized divergence incorporating nonlinearity, and depends on the solution \cite{Efrati2009,Moshe2014ISF,Moshe2015PRE}. 

Despite its complications, \Eqref{eq:div} can be analyzed by adopting a geometric approach to elasticity.
In this approach the basic quantity is the \emph{metric} denoted $\g$, a  $2^\text{nd}$ rank tensor quantifying distances between neighboring elements. For example, the distance between points separated in coordinates by a vector $\dif x^\mu$ is $\dif l^2 = \g_{\mu\nu} \dif x^\mu\dif x^\nu$. 
Upon discarding the notion of displacement field, two metrics are defined on a solid, the target and actual metrics $\gbar$ and $\g$ locally quantifying rest and actual distances between neighboring material elements. For a homogeneous solid with no internal structure the target metric in Cartesian coordinates can be set to $\gbar_{ij} = \delta_{ij}$.

In this approach the strain is defined
\begin{equation}
\us = \frac{1}{2}\brk{\g - \gbar}
\label{eq:GeometricStrain}
\end{equation}
When $\g$ is derived from a displacement field, both definitions \eqref{eq:Strain} and \eqref{eq:GeometricStrain}  coincides, confirming that the metric formulation is still geometrically nonlinear.

A direct analytical approach to solve \eqref{eq:div} is by representing the solution in terms of a single scalar potential $\chi$, which is known as Airy Stress Function in the case of fully linearized elasticity \cite{LandauElasticityBook}. This approach is very similar to the electrostatic problem, where one of Maxwell's equations is identically solved by representing the electric field as the gradient of a potential. Recently it was shown that a generalization for the representation of $\bfsigma$ in the nonlinear theory exists, and $\chi$ is termed  \emph{Incompatible Stress Function} \cite{Moshe2014ISF,Moshe2015PRE}.
While the representation might seem complicated and not very informative, we present it here to emphasize a simple observation
\begin{eqnarray}
\bfsigma =  \nabla^{\g} \times { \chi \times \nabla^{\gbar} }.
\label{eq:ISF}
\end{eqnarray}

Since the unknown is the actual metric $\g$, the above representation is implicit, as $\g$ appears both in the left hand side through the relation of stress with strain in \eqref{eq:Hook}, and in the right hand side through the covariant derivative $\nabla^{\g}$.

While \Eqref{eq:ISF} identically solves \Eqref{eq:div}, an additional geometric constraint on $\g$ is required to enforce a flat configuration.  Unfortunately, the implicit form of the representation \eqref{eq:ISF} hinders further analytical progress.
However, in cases where a dimensionless small parameter $\eta$ exists, a progress can be made by representing $\g$ and $\chi$ as series expansions: 
\begin{equation}
\begin{split}
\g &= \gbar + \eta \, \delta \g^{1st} + \eta^2 \, \delta \g^{2nd} + \dots \\
\chi &=  \eta \, \chi^{1st} + \eta^2 \, \chi^{2nd} + \dots \\
\end{split}
\label{eq:FormalExpansion}
\end{equation}
In the case of a simple perforated solid, requiring a flat metric $\g$ results with a sequential set of equations 
\begin{equation}
\begin{split}
\Deltabar \Deltabar \chi^\text{1st} & = 0\\
\Deltabar \Deltabar \chi^\text{2nd} & = F^\text{2nd} \brk{\chi^\text{1st}}\\
\Deltabar \Deltabar \chi^\text{3rd} & = F^\text{3rd} \brk{\chi^\text{1st},\chi^\text{2nd}}\\
&\vdots
\end{split}
\label{eq:GeneralizedBiHarmonic}
\end{equation}
where $\Deltabar$ is the Laplace-Beltrami operator with respect to the reference metric $\gbar$, and the functions $F$ on the right-hand sides are found by a direct computation, making each order depending on previous solutions. 

Very similar to the electrostatic analog, incorporating image elastic charges into the theory can be easily done by including source term in the right-hand side of the first equation in \eqref{eq:GeneralizedBiHarmonic}. The reader that is familiar with this formalism may wonder why the formation of image charge is not reflected in a re-definition of the reference metric. This is simply because the internal structure of the material is not modified in response to formation of image elastic charge, implying these are singularities of the actual metric $\g$, and not of the reference one.
 
By plugging $q^{\alpha\beta} \bar{\nabla}_{\alpha\beta} \delta\brk{\xvec}$ in the right hand side of the first equation in \eqref{eq:GeneralizedBiHarmonic} we find, for example, the stress functions associated with a quadrupole. In the attached Mathematica notebook we derive close analytic forms for the stress functions of different elastic multipoles up to fourth order. While this calculation is mostly technical, this is a central part of the work that allowed us to go beyond the linear elastic model and obtain explicit expression for $\M$ and $\N$ in  \eqref{eq:QuadEnergy}.

\paragraph*{Example: Isotropic mode of quadrupole nonlinear solution}
As an example we now present the nonlinear solution to the problem of an isolated isotropic local quadrupole charge, describe by the equation
\begin{equation}
\frac{1}{Y} = \Delta \Delta \chi^{1st} = p \Delta \delta\brk{\xvec}
\end{equation}
Here $p \Delta \delta\brk{\xvec}$ is the Laplacian of Dirac delta function \cite{Moshe2015PRE}.
A detailed derivation of this solution and others are given in an attached Mathematica notebook. The nonlinear stress function is 
\begin{equation}
\chi =  p \, \chi^{1st} + p^2 \, \chi^{2nd} + \dots 
\end{equation}
with
\begin{equation}
\begin{split}
\chi^{1st} &= \frac{Y}{2} \log \brk{r}\\
\chi^{2nd} &= -\frac{Y \brk{1+\nu^2}}{2\, r^2}\\
\chi^{3rd} &= -\frac{Y (1+\nu)^2 \left(\nu^2-13\right)}{3 \, r^4}\\
\chi^{4th} &= -\frac{(1+\nu)^4 \left(7 \nu ^2-37 \nu +26\right) Y}{18 r^6}
\end{split}
\end{equation}
From this stress function expansion we derive the different orders of the displacement, stress, and strain distributions.

\section{Two types of multipoles}
\label{app:Hex}  
Elastic charges are quantified by curvature singularity of a reference geometry \cite{Seung88,Moshe2015PNAS}.  Monopole charges correspond to delta-function singularity, leading to an equation of the form
\begin{equation}
\frac{1}{Y} \Delta\Delta \chi^\text{1st} = m \,\delta\brk{\xvec}
\end{equation}
The monopole is topologically protected and therefore is a conserved quantity \cite{Kupferman2013ARMA}. There is another isotropic charge that is not conserved, but corresponds to local elastic deformations, that is the isotropic Eshelby inclusion \cite{Eshelby1957}. Within the formalism of elastic charges this object is quantified as the Laplacian of delta function singularity \cite{Moshe2015PNAS}
\begin{equation}
\frac{1}{Y} \Delta\Delta \chi^\text{1st} = p \, \Delta \delta\brk{\xvec}
\end{equation}
Below, the delta function singularity will be called a topological monopole, and the Laplacian of delta function a non-topological monopole. 
Multipole elastic charges are quantified by derivatives of the monopole singularity. For example, a (pure) quadrupole of the first type is composed of 4 topological monopole charges of alternate signs and described by charge singularity of the form
\begin{equation}
Q^{\alpha\beta} \nabla_{\alpha\beta} \delta\brk{\xvec}
\end{equation}
with $Q$ a traceless symmetric tensor. Similarly, a quadrupole of the second kind corresponds to four non-topological monopole charges of alternate signs, and described  by charge singularity of the form
\begin{equation}
H^{\alpha\beta} \nabla_{\alpha\beta} \Delta \delta\brk{\xvec}
\end{equation}
with $H$ a traceless symmetric tensor.

All elastic multipoles can be described as multipoles of the first and second kind as described above.

In the present work we allow for the formation of image quadrupoles inside each hole, hence we have a total of four degrees of freedom (2 in $Q$ and 2 in $H$). In addition, we allow for the formation of non-topological monopole, leading to a total of 5 degrees of freedom in each hole. 

In the attached Mathematica notebook we calculate the nonlinear elastic field of these 5 modes.

\section{Variational formulation of electrostatics with conductive boundary conditions}
\label{app:VariationalApproach}
In this appendix we derive the variational formulation of the electrostatic problem described in \eqref{eq:ElecFE}.

Consider an electrostatic potential defined on the domain $\Omega$ with conductive boundary conditions on $\partial \Omega$. The electric potential satisfies
\begin{equation}
    \begin{split}
        \Delta \phi \brk{\xvec} & = \rho\brk{\xvec} \quad \forall\, \xvec \in \Omega \\
        \vec\nabla \phi \cdot \hat{t} & = 0  \quad \forall \, \xvec \in \partial\Omega 
\end{split}
\end{equation}
We show that this boundary value problem is equivalent to variational minimization of the following functional
\begin{equation}
    F = \int \frac12 \brk{\vec \nabla \phi \cdot \vec \nabla \phi}  \dif S - \oint  f\, \vec \nabla \phi  \cdot  \dif \vec l
    \label{eq:Energy1}
\end{equation}
where the variation is performed with respect to the electrostatic potential $\phi$ and the local Lagrange multiplier $f$.
Indeed, variation with respect to $f$ yields
\begin{equation}
    \delta F_{f} =  - \oint  \delta f\, \vec \nabla \phi  \cdot  \dif \vec l 
\end{equation}
Since at minimum this expression vanishes for any $\delta f$, we find
\begin{equation}
    \vec \nabla \phi \cdot \hat{t} = 0 \quad \,\forall \xvec \in \partial \Omega
\end{equation}
where $\hat{t}$ is the tangent to the boundary. This condition corresponds to a constant potential on the boundary, which in turns describes a conductive boundary condition.

Next, the variation with respect to $\phi$ 
\begin{equation}
    \delta F_{\phi} = \int_{\Omega} \vec \nabla \phi  \cdot \vec \nabla {\delta \phi}  \dif S - \oint_{\partial \Omega}  f\, \vec \nabla \delta \phi  \cdot  \dif \vec l
\end{equation}
Upon integrating by parts and using the identity $\oint (h \vec \nabla g - g \vec \nabla h) \, \dif \vec{l} = 0$ we find
\begin{equation}
    \delta F_{\phi} = \oint_{\partial \Omega} \brk{\vec \nabla \phi  \cdot \hat{n}} {\delta \phi}  \, \dif l - \int_{\Omega} \Delta \phi \, {\delta \phi}  \, \dif S - \oint  \delta \phi \, \vec \nabla f  \cdot  \dif \vec l
\end{equation}
Since at minimum this expression vanishes for any $\delta \phi$ we find
\begin{equation}
    \begin{split}
        \Delta \phi &= 0 \quad \forall \xvec \in \Omega \\
        \vec\nabla f \cdot \hat{t} &= \vec \nabla \phi \cdot \hat{n} = E_\perp \quad \forall \xvec \in \partial \Omega 
    \end{split}
\end{equation}
Note that on a conductive surface the normal electric field is equal to the charge density $E_\perp = \rho$. 

\section{Finding the optimal charges}
\label{app:NLImplementation}

\newcommand{\Q}{\bm{Q}}
\newcommand{\D}{\bm{D}}
\renewcommand{\H}{\bm{H}}
\renewcommand{\P}{\bm{P}}
\newcommand{\W}{\bm{W}}
\newcommand{\C}{\bm{C}}
\newcommand{\bl}{\bm{\lambda}}
\newcommand{\ex}[1]{^{(#1)}}
\newcommand{\pa}{\partial}
As explained in the main text, the optimal charges are the ones that by minimize the energy functional
\begin{align}
U&=\sum_{ij} Q_i Q_j M_{ij}  \ex{1}+\sum_{ij} 2Q_i^2 Q_j M\ex{2}_{ij} + \mathcal{O}(\Q^4)
\label{eq:app_energy}
\end{align}
under the nonlinear constraints function
\begin{align}
  \nonumber
  C_j(\Q)&\equiv-D_j + 
 \sum_i Q_i\N_{ij}\ex1+ Q_i^2 \N_{ij}\ex2 + Q_i^3\N_{ij}\ex3
 \\&=0+ \mathcal{O}\left(\Q^4\right)\ .
 \label{eq:app_constraint}
\end{align}
$\C$ is a vector of constraints of length $c$, one constraint per ligament. The dimensions of these matrices are
\begin{align}
\nonumber
\Q &\in  \mathbb{R}^n\ , &
\C, \D &\in  \mathbb{R}^c\ , &
\M &\in  \mathbb{R}^{n,n}\ , &
\N &\in  \mathbb{R}^{n,c}\ .
\end{align}
where $n$ is the number of degrees of freedom, i.e.~5 per hole (three for one type of quadrupole, two for the other, see \appref{app:Hex}). To impose the constraints, we introduce a vector $\bl\in\mathbb{R}^c$  whose components are Lagrange multipliers. Then, the function to minimize is 
\begin{align}
    F=U(\Q)-\bl\cdot\C(\Q)
    \label{eq:app_functional}
\end{align}

To make the expansion in small displacement, we formally write the constraint as $\D=z \tilde{\D}$ where $z$ is a small parameter and $\tilde{\D}$ is some fixed vector (for uniform displacement $\D$ is a constant vector of ones).

The leading order solution is obtained by neglecting the cubic term in Eq.~\eqref{eq:app_energy} and the quadratic term in Eq.~\eqref{eq:app_constraint}. The gradient of Eq.~\eqref{eq:app_functional} is then $2\M\Q- \N\bl$
and it vanishes for $\Q_*\equiv\frac{1}{2}\M^{-1}\N\bl$. To find the actual solution, we need to solve for $\bl$ by demanding $\C=0$, yielding
\begin{align}
\N ^T\Q_* &= z\tilde{\D}\ , &
\bl&=2z\left(\N^T\M^{-1}\N\right)^{-1}\tilde{\D}
\label{eq:app_linear_solution}
\end{align}
Finally, we obtain the solution for $\Q_*$. We write $\Q_*=z \Q_0$ with
\begin{align}
\Q_0 = \M^{-1}\N\left(\N^T\M^{-1}\N\right)^{-1}\tilde{\D} \ .
\end{align}

\subsection{Stability (nonlinear) analysis}
For the next order calculation we are not concerned with finding the minimum of the cubic functional, which is technically involved, but only with the stability of the leading order (linear) solution. To this end, we look for perturbations around it, i.e. $\Q=z\Q_0+\delta\Q$.

The hessian at $\Q_0$ is
\begin{widetext}
\begin{align}
\H = \left. \frac{\pa^2 F}{\pa Q_i \pa Q_j} \right|_{\bm Q_0}& =
2M_{ij}\ex 1+
4z\underbrace{\left[
\diag{}(\Q_0) \M\ex{2}
+\left(\diag(\Q_0) \M\ex{2}\right)^T
+\diag\left(\M\ex2 \Q_0\right)
\right]}_{=d\H}
+\mathcal{O}(z^2)
\end{align}
Note the definition of $d\H$, the linear variation in the Hessian with $z$. Expanding $\C$ in powers of $z$, we get
\begin{align}
  \C(z \Q_0 + \delta \Q)
  &= -z\D+(z\Q_0+\delta \Q)^T\N\ex{1}+\left((z\Q_0+\delta \Q)^2\right)^T\N\ex{2}+\cdots \\
  &= \delta\Q^T
  \Big[\N\ex{1}+2z\underbrace{\diag(\Q_0)\N\ex{2}}_{=d\N}\Big]
  +\mathcal{O}\left(z^2, \delta\Q^2\right)
\end{align}
\end{widetext}
where $\diag(\Q_0)$ is a diagonal $n\times n$ matrix whose diagonal is $\Q_0$ and we used the fact that $\Q_0^T\N\ex{1}=\D$ as guaranteed by Eq.~\eqref{eq:app_linear_solution}. Note the definition of $d\N$, the linear variation in $\nabla \C$ with $z$.

To impose that  $\delta\Q$ does not violate the constraints we demand that
\begin{align}
  \delta\Q^T(\N\ex{1}+z d\N)=0\ .
\end{align}
This is a linear constraint on $\delta\Q$, which we can satisfy identically by writing $\delta \Q = \bm P(z) \bm q$, where $\bm P(z)$ is a projection matrix of size $n\times (n-c)$, $\bm q$ is the reduced-dimensional coordinate and $\bm P(z)$ satisfies
$\bm P(z)^T(\N\ex{1}+z d\N)=0$.

The Hessian restricted to the allowed directions is $\bm P^T\bm H\bm P$ where $\bm H$ is the original Hessian. The critical point is calculating by finding the smallest $z$ for which the constrained Hessian has a zero eigenvalue. That is, we find the minimal $z$ such that an eigenvalue of $\P(z)^T(\M\ex{1}+z d\H)\P(z)$ vanishes.

\bibliography{references}  
\end{document}